\documentclass[aps,pra,reprint]{revtex4-2}
\usepackage{graphicx}
\usepackage{dcolumn}
\usepackage{bm}
\usepackage{float}
\usepackage{amssymb}
\usepackage{xcolor,soul} 
\usepackage{braket}
\usepackage{hyperref}
\usepackage[utf8]{inputenc}
\usepackage[T1]{fontenc}

\newcommand\Ca{$^{40} \text{Ca}^+$}

\begin{document}
\title{Efficient Three-Dimensional Sub-Doppler Cooling of \Ca~in a Penning Trap}

\author{Brian J. McMahon}
\email{brian.mcmahon@gtri.gatech.edu}

\author{Brian C. Sawyer}%
\affiliation{Georgia Tech Research Institute, Atlanta, GA 30332, USA}%

\date{\today}
\begin{abstract}
We demonstrate efficient sub-Doppler laser cooling of the three eigenmodes of a \Ca~ion confined in a compact Penning trap operating with a magnetic field of 0.91~T. Using the same set of laser beams as required for the initial Doppler laser cooling operation, we detune the laser frequencies to produce a narrow two-photon dark resonance. The process achieves a $1/e$ cooling time constant of $108(8)~\mu\text{s}$, ultimately reducing the mean thermal axial mode occupation from $72(23)$ to $1.5(3)$ in $800~\mu$s as measured by resonantly probing an electric quadrupole transition near 729~nm. A parametric drive is applied to the trap electrodes which coherently exchanges the axial mode occupation with that of each radial mode, allowing for three-dimensional sub-Doppler cooling using only the axially-propagating laser beams. This sub-Doppler cooling is achieved for an axial oscillation frequency of $\omega_z=2\pi\times221$~kHz, which places the motion well outside of the Lamb Dicke confinement regime at the Doppler laser cooling limit. Our measured cooling rate and final mode occupation are in good agreement with a semiclassical model which combines a Lindblad master equation solution for ion-photon interactions with classical harmonic oscillator motion of the trapped ion. 
\end{abstract}

\maketitle 

\section{Introduction}
Arrays of atomic ions confined in surface electrode Penning traps are identified as a potential architecture for quantum information processing~\cite{jain_penning_2024}. Finite phonon occupation of motional modes of trapped ions can degrade the fidelity of entangling gate operations, with the highest-fidelity laser-based entanglement demonstrations employing ground state cooling~\cite{gaebler_high-fidelity_2016, ballance_high-fidelity_2016, clark_high-fidelity_2021, baldwin_high-fidelity_2021}. While sub-Doppler laser cooling of single ions and Coulomb crystals in radiofrequency (rf) ion traps is now commonplace (e.g.~\cite{roos_experimental_2000, lechner_electromagnetically-induced-transparency_2016,allcock_dark-resonance_2016,scharnhorst_experimental_2018,feng_efficient_2020,wu_electromagnetically-induced-transparency_2025,bartolotta_laser_2024, huang_electromagnetically_2024}), demonstrations with ions in Penning traps have thus far been less numerous~\cite{jordan_near_2019,hrmo_sideband_2019,goodwin_resolved-sideband_2016,jain_penning_2024,cornejo_resolved-sideband_2024}. Resolved sideband cooling has been applied to small $^{40}$Ca$^+$ ion arrays and single $^9$Be$^+$ ions \cite{jain_penning_2024,cornejo_resolved-sideband_2024}, while electromagnetically-induced transparency (EIT) cooling in a Penning trap has been demonstrated using two-dimensional crystals of $^9$Be$^+$~\cite{jordan_near_2019}. 

Singly-ionized calcium is an attractive choice of ion species for quantum operations in Penning traps because of its relatively low mass and convenient laser wavelengths for electronic state manipulation compatible with integrated photonics~\cite{mordini_multizone_2025,xing_rapid_2025}. While sideband cooling is an effective method for cooling atomic motion to the ground state, in the regime of strong confinement (Lamb Dicke parameter, $\eta\lesssim0.1$), the cooling speed is limited by the reduced sideband interaction strengths proportional to $\eta$. Conversely, in the weak confinement regime ($\eta \gtrsim 0.1$), beyond-first-order sidebands must be driven in order to ground state cool, which substantially extends the total cooling time~\cite{roghani_trapped-atom_2008,goodwin_resolved-sideband_2016,hrmo_sideband_2019}.

In this article, we demonstrate efficient dark resonance (DR) cooling of the axial mode of a single \Ca, reducing the initial Doppler-cooled thermal occupation from $>70$ to $<2$ quanta in 800~$\mu$s. From this pre-cooled temperature, a few rounds of pulsed sideband cooling reduce the remaining motional occupation to $<1$. In comparison with resolved sideband cooling in a comparable \Ca~Penning trap system, our combination of DR and sideband cooling reduces the axial ground state cooling time from 20~ms~\cite{goodwin_resolved-sideband_2016} to 3.8~ms. To efficiently cool the radial degrees of freedom, the ground-state-cooled axial mode population is coherently exchanged with that of a Doppler-cooled radial mode. This is achieved by applying an oscillating quadrupolar potential to the trap electrodes that is resonant with the sum or difference of the two coupled mode frequencies \cite{cornell_mode_1990,gorman_two-mode_2014}. By applying sub-Doppler cooling along only the axial trap direction and employing this motional mode exchange, we efficiently sub-Doppler cool all degrees of freedom.

Penning trap ion confinement is achieved through the combination of a large, uniform static magnetic field and an electrostatic quadrupole. Ion motion parallel to the magnetic field is harmonic with an oscillation frequency $(\omega_z)$ determined solely by the electric quadrupole potential. Motion perpendicular to the magnetic field with magnitude, $B$, is governed by a combination of the Lorentz force and the electrostatic force from the quadrupole potential. This results in two radial eigenmodes, the modified cyclotron mode ($\omega_+$) and the magnetron mode ($\omega_-$) with frequencies~\cite{itano_laser_1982}
\begin{equation}
\omega_{\pm} = \frac{\omega_c}{2} \pm \frac{\sqrt{\omega_c^2-2\omega_z^2}}{2},
\end{equation}
where $\omega_c = qB/m$ is the bare cyclotron frequency for an ion with mass, $m$, and charge, $q$. At the 0.91~T trapping magnetic field for the apparatus described in this article, we achieve $\omega_c =  2\pi\times351$~kHz, which requires $\omega_z < \omega_c / \sqrt{2} =  2\pi\times 248$~kHz for stable radial confinement. Our trap frequencies for this work are $(\omega_z,\omega_+,\omega_-)=2\pi\times (221~\text{kHz}, 256~\text{kHz}, 96~\text{kHz})$.

Doppler laser cooling of single ions and Coulomb crystals in Penning traps has been well-characterized both theoretically and experimentally~\cite{itano_laser_1982,jensen_temperature_2004,koo_doppler_2004}. The axial ion motion experiences no Lorentz force in an ideal Penning trap, therefore axial Doppler cooling dynamics are identical to that of rf traps. Doppler cooling of the modified cyclotron mode is similarly straightforward, while the occupation of the negative-energy magnetron mode can only be reduced through heating. This mismatch in radial mode cooling requirements complicates three-dimensional Doppler cooling in Penning traps, and is typically resolved using a radially-displaced Doppler cooling laser beam~\cite{wineland_laser_1979} and/or application of an oscillating quadrupolar potential that couples the two radial modes (i.e. axialization)~\cite{powell_improvement_2003}. We employ both radial cooling techniques simultaneously for this work. The Doppler cooling limit of $\sim0.5$~mK implies an average motional occupation of $\sim40-100$ for our trap frequencies. To our knowledge, sub-Doppler laser cooling of the magnetron motion of a trapped ion has thus far only been demonstrated using resolved sideband cooling~\cite{hrmo_sideband_2019, jain_penning_2024}. 

The remainder of this article is structured as follows: Sec.~\ref{sec:experiment} describes our experimental setup and demonstration of DR and sideband cooling combined with resonant mode exchange, while Sec.~\ref{sec:model} details our semiclassical DR cooling model relevant to the regime of weak ion confinement.

\section{Experimental Setup} \label{sec:experiment}
All experiments in this article are performed with single \Ca~ions trapped in a compact permanent magnet Penning trap. The trap is identical to that described in Ref.~\cite{mcmahon_individual-ion_2024}. Briefly, a pair of radially-magnetized SmCo magnet assemblies provides the uniform trapping magnetic field with a strength of $\sim0.91$~T. Two parallel printed circuit boards (PCBs) with mirrored electrode patterns provide the static electric fields for harmonic confinement along the magnetic field direction. Azimuthally segmented electrodes on the PCBs are employed to create an ac quadrupolar potential for mixing the energy of the two radial modes for axialization during Doppler cooling~\cite{hendricks_laser_2008,savard_new_1991,powell_axialization_2002,guan_broadband_1994,phillips_dynamics_2008,powell_improvement_2003}). These electrodes are also used for coherent motional population exchanges between each radial mode and the axial mode \cite{cornell_mode_1990}.

\begin{figure}
    \centering
    \includegraphics[scale=0.9]{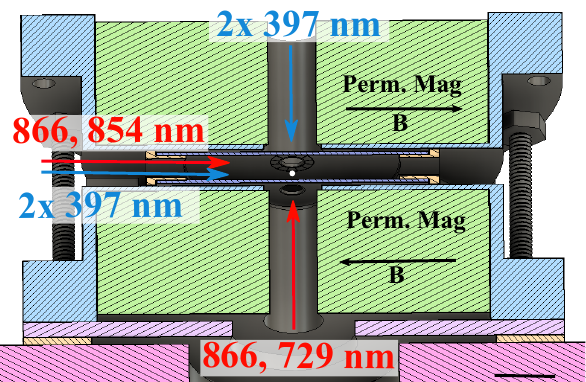}
\caption{Cross-section view of the in-vacuum permanent magnet Penning trap assembly with printed circuit boards. The pair of radially magnetized SmCo rings produces a uniform, vertically-oriented magnetic field of 0.91~T at the ion location. The ion is depicted as a white circle centered vertically between the PCBs. The laser propagation directions are shown as blue and red arrows. The scale bar (bottom right) represents 12.7~mm for this drawing.}
    \label{fig:laserbeamdirections}
\end{figure}

The trapped \Ca~motion is cooled with two distinct sets of laser beams. The Doppler cooling beams enter the vacuum chamber via two photonic crystal fibers along the axial and radial directions as shown in Fig.~\ref{fig:laserbeamdirections}. The radial beam path consists of two 397~nm laser beams (397A and 397B) co-propagating with an 866~nm and an 854~nm beam. The axial cooling laser beam path consists of two co-propagating 397~nm laser beams (397A and 397B) and a counter-propagating 866~nm beam. Due to the large Zeeman shifts of magnetic sublevels in the Penning trap, the 866~nm and 854~nm beams are individually modulated with fiber electro-optic phase modulators (EOMs) to efficiently clear population from all metastable $D_{3/2}$ and $D_{5/2}$ states~\cite{koo_doppler_2004}. The relevant optical transitions and \Ca~level structure are illustrated in Fig.~\ref{fig:leveldiagram}. We characterize axial ion temperature using a 729~nm laser beam that co-propagates with the axial 866~nm laser beam and excites the narrow-linewidth $S_{1/2},\ m_J=-1/2\rightarrow D_{5/2},\ m_J=-3/2$ transition, where $m_J$ is a magnetic sublevel of a state with total angular momentum, $J$. Every experiment starts with Doppler cooling which reduces the motional mode temperatures to near their respective Doppler limits.

\subsection{Axial Dark Resonance Cooling}

\begin{figure}
    \centering
    \includegraphics[scale=0.75]{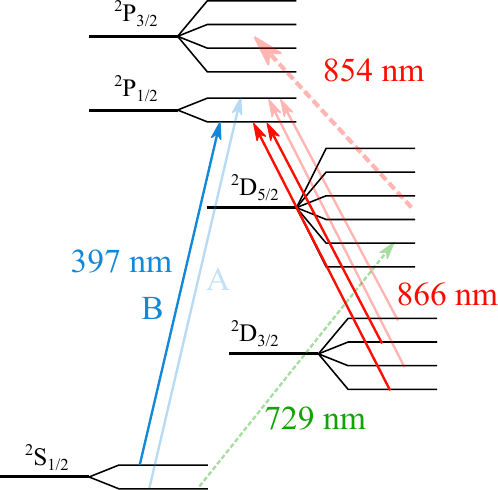}
    \caption{The \Ca~energy level diagram with relevant laser wavelengths used for this work. The solid and transparent arrows denote laser frequencies used for Doppler and dark resonance cooling, with the distinct sets identifying excitation pathways to each of the $P_{1/2}$ magnetic sublevels. The 729~nm excitation (green dashed arrow) is used to measure axial mode occupation, and a set of 854~nm tones (red dashed arrow) returns population to the $S_{1/2}$ state following 729~nm excitation.}
    \label{fig:leveldiagram}
\end{figure}

Reducing ion temperature to below the Doppler limit requires a distinct cooling scheme. Dark resonance cooling utilizes off-resonant excitation via a pair of laser frequencies that creates a narrow two-photon (i.e. dark) resonance. This dark resonance is useful for sub-Doppler cooling due to its tuneable bandwidth that can be made smaller than that of the $S\rightarrow P$ resonance feature. At low magnetic field, the two DR cooling laser frequencies can be conveniently produced via a single laser source and distinct acousto-optic modulators (AOM)~\cite{roos_experimental_2000}. However, at Tesla-scale magnetic fields, the relatively large Zeeman splittings necessitate a different approach (i.e. distinct phase-locked lasers as in Ref.~\cite{jordan_near_2019}). 

In our system, we have chosen to employ the same laser beams as used for Doppler cooling, but with different powers and detunings for the DR cooling step. The two lasers creating the DR must have sufficient relative frequency stability to maintain a fixed two-photon detuning throughout the DR cooling process. To achieve this, we stabilize one of the 397~nm lasers (397B) and the 866~nm laser (both extended cavity diode lasers) to a common temperature-stabilized, high-finesse ($\mathcal{F} \sim 3000$ for both wavelengths) optical cavity with a specified drift of $<2$~kHz/hour. The other 397~nm laser (397A) and the 854~nm laser are locked with $<2$~MHz instability to a commercial wavelength meter. The wavelength meter is recalibrated hourly to a reference 852~nm laser which is locked via saturated absorption spectroscopy to a Cs D2 transition in a room-temperature vapor cell.

\begin{figure}
    \centering
    \includegraphics[scale=0.7]{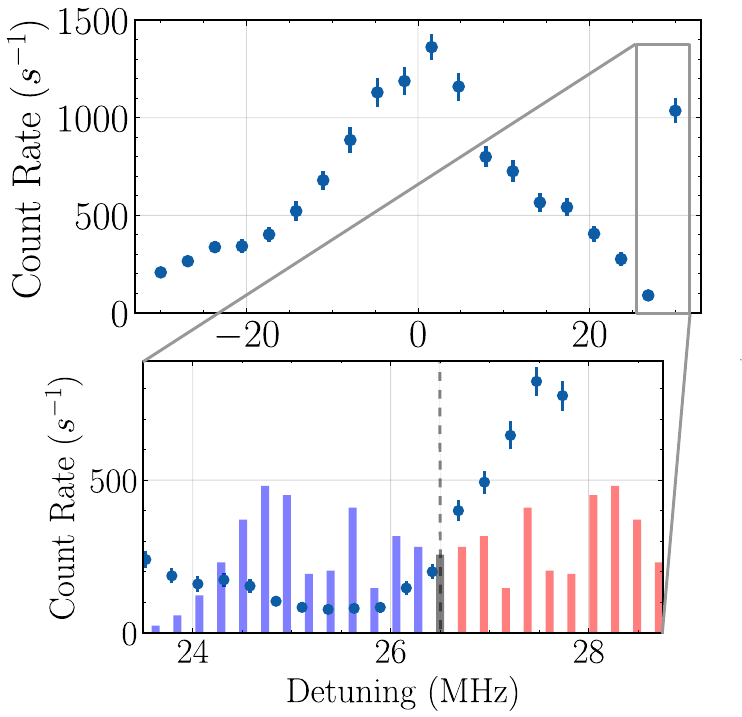}
    \caption{(Top) Observed ion fluorescence count rate measured for various detunings of the 397B laser from atomic resonance. (Bottom) Zoomed-in view of the two-photon resonance from the frequency scan above. The transparent grey, dashed line shows the detuning of the 397B laser frequency (i.e. optical carrier) during DR cooling. The frequency offsets and relative Rabi rates of the red and blue axial sidebands are plotted on an arbitrary vertical scale for an initial coherent axial occupation of $\bar{n}=72$. Net axial cooling occurs when the photon scatter rate for energy-removing sidebands overwhelms that of the energy-adding sidebands.}
    \label{fig:GTRI_DRC}
\end{figure}

The axially-oriented 397~nm laser beams are independently controlled via AOMs aligned in a double-pass configuration, while the axially-oriented 866~nm laser beam is independently controlled with a fiber AOM. For DR cooling, the modulators for the 397B and 866~nm lasers are each shifted to the blue of the Doppler cooling resonance by $\sim 26$~MHz to create a two-photon DR. The DR is measured by scanning the frequency of the stabilized 397B laser as shown in Fig.~\ref{fig:GTRI_DRC}. The resonance is formed between the 397B laser which connects $S_{1/2}$, $m_J=1/2$ to $P_{1/2}$, $m_J=-1/2$ and one of the 866 nm laser EOM tones which connects a $D_{3/2}$ state (either $m_J=-3/2$ or $m_J=1/2$) to the same $P_{1/2}$, $m_J=-1/2$ state. These particular transitions are highlighted in Fig.~\ref{fig:leveldiagram} as the bold arrows. The optical powers and detunings of each beam are optimized for the steepest slope on the energy removing side of the two-photon resonance while also achieving the least excitation on the energy adding side. Typical experiment parameters for the data of Fig.~\ref{fig:GTRI_DRC} include 1~$\mu$W and 6~$\mu$W for the 397B and 397A laser beam powers, respectively, and 500~$\mu$W total 866~nm laser beam power. The laser beam waists at the ion position are $\sim140$~$\mu$m and $\sim200$~$\mu$m for the 397~nm and 866~nm beams, respectively. The corresponding Rabi rates are $\Omega_{397A}\sim 2\pi\times 2.9$~MHz, $\Omega_{397B}\sim2\pi\times 1.2$~MHz, and $\Omega_{866}\sim 2\pi\times (1.3 - 2.2)$~MHz for the 397A, 397B, and 866~nm excitations. These parameters yield a two-photon resonance feature with $\sim2$~MHz full-width at half maximum (see inset of Fig.~\ref{fig:GTRI_DRC}). We note that this sub-Doppler cooling regime is similar to that of Ref.~\cite{jordan_near_2019} with nearly equal Rabi rates, and contrasts with canonical EIT cooling schemes which employ very asymmetric Rabi rates between designated `pump' and `probe' excitations~\cite{morigi_cooling_2003}. 

\begin{figure}
    \centering
    \includegraphics[scale=0.31]{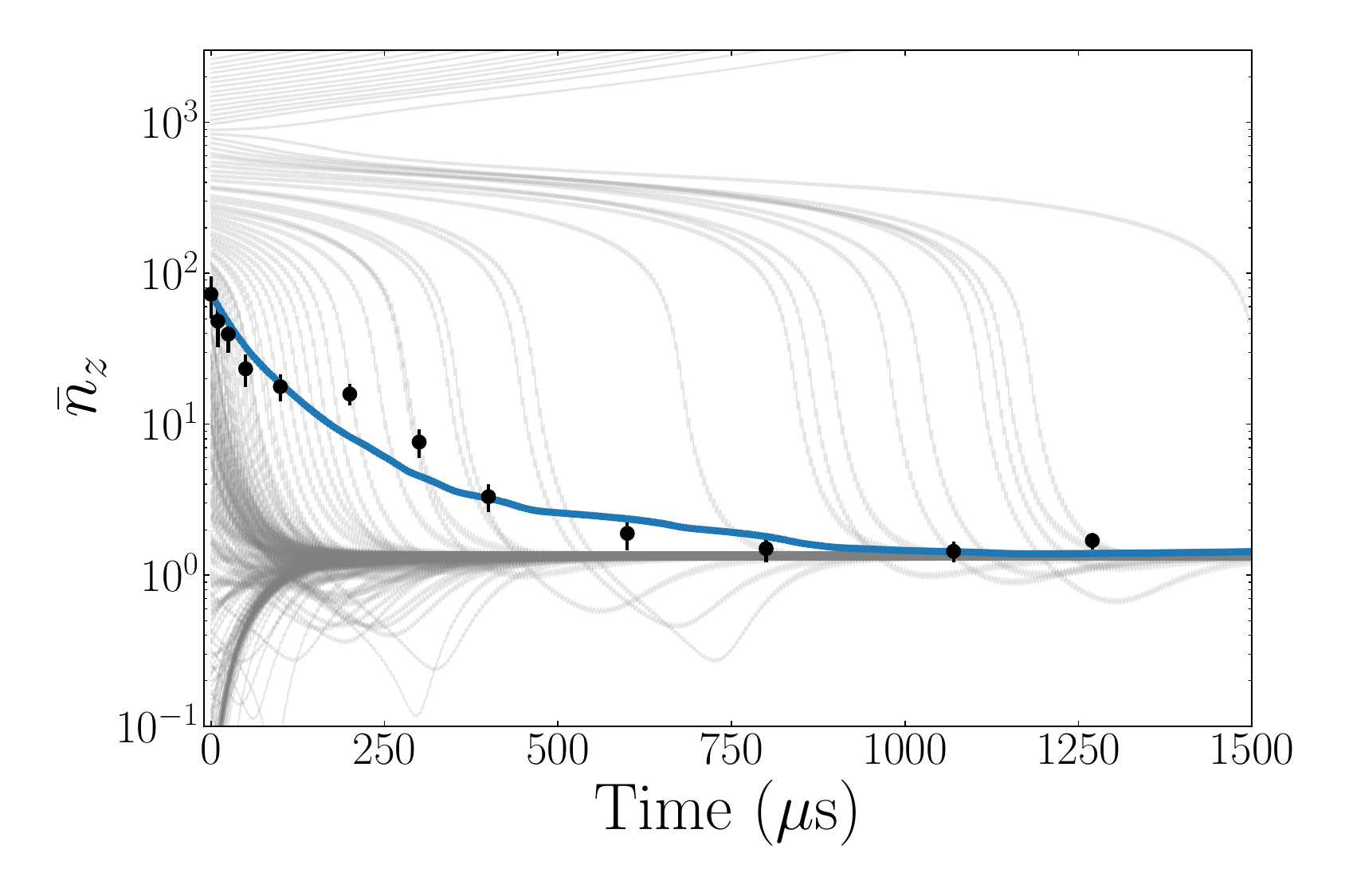}
    \caption{Measured axial mode occupation (points with error bars) of \Ca~after application of DR cooling for various durations. The solid blue line is the result of a semiclassical sub-Doppler cooling simulation that takes experimental parameters as inputs. Solid gray lines show the individual cooling (or heating) trajectories for each simulated initial coherent excitation and the blue line is a thermally-weighted mean over all trajectories. We find good agreement between experiment and simulation over nearly two orders of magnitude in axial mode occupation. The simulated DR cooling capture range is $\bar{n}_z < 900$ for these parameters, which include Rabi rates $\Omega_{397A}\sim2\pi\times 2.9$~MHz, $\Omega_{397B}\sim2\pi\times 1.2$~MHz, and $\Omega_{866}\sim 2\pi\times(1.3-2.2)$~MHz.}
    \label{fig:drcvstime}
\end{figure}

Our axial frequency of $\omega_z=2\pi\times221$~kHz combined with the counter-propagating 397~nm and 866~nm laser beam geometry yields an effective Lamb Dicke parameter of $\eta_z\sim0.55$ for DR cooling. We typically realize the \Ca~Doppler cooling limit of $\sim0.5$~mK in 5-10~ms, which leads to $\eta_z\sqrt{2\bar{n}_z+1}>5$, placing the ion motion well outside of the Lamb Dicke regime at the start of DR cooling. Sidebands generated by the ion oscillation along the DR laser beam propagation direction have non-negligible Rabi rates beyond $\pm 2$~MHz (see inset of Fig.~\ref{fig:GTRI_DRC}). The initial 397B laser detuning for DR cooling is adjusted for the largest differential excitation (measured via photon scatter) between red and blue sidebands. The final optimization of the 397B laser detuning is completed by measuring the axial temperature as a function of DR cooling duration. As shown in Fig.~\ref{fig:drcvstime}, we characterize the DR cooling rate and final axial mode occupation via fits to 729~nm carrier flops.

From Fig.~\ref{fig:leveldiagram}, the two distinct 397~nm laser excitations allow for two sets of dark resonance conditions that may be tuned independently of one another. The coldest ion temperatures are measured when positioning the 397A laser near its respective two-photon resonance (shown as the transparent arrows in Fig. \ref{fig:GTRI_DRC}). Figure~\ref{fig:drcvstime} demonstrates this with a final axial mode occupation of $\bar{n}_z=1.5(3)$ in $800~\mu$s from an initial Doppler-cooled occupation of $\bar{n}_z=72(23)$. A single-exponential fit to the data points of Fig.~\ref{fig:drcvstime} (not shown) yields a DR cooling time constant of $108(8)~\mu$s.

\subsection{Parametric Mode Coupling}
Parametric mode coupling can be used to coherently swap the population between motional modes. In our system, this is performed by applying the appropriate sum or difference frequency to the azimuthally segmented ring electrodes on the PCBs. This creates an oscillating quadrupolar potential tilted with respect to the magnetic field which can couple the axial and radial motion~\footnote{Our trap PCB electrodes produce an electrostatic quadrupole that is slightly tilted with respect to the trap magnetic field. We remove this tilt with a waveform that differentially charges the azimuthally-segmented electrodes. This means that a uniform oscillating voltage applied to all segments will introduce an oscillating quadrupolar potential that is tilted with respect to the magnetic field.}. The axial and magnetron modes are coupled by applying the sum frequency of the two modes $\omega_z + \omega_-$, while the modified cyclotron motion is coupled to the axial mode via application of the difference frequency, $\omega_+ - \omega_z$ \cite{cornell_mode_1990}. Before calibrating the mode coupling, the motional mode frequencies are first measured by scanning the rf frequency of a pulse applied to the same electrodes. When the rf pulse is resonant with the frequency of a particular mode, the energy of the mode grows with the applied pulse duration until the temperature increase can be seen as a reduction of ion fluorescence. Typically, the mode frequencies are calibrated once per day with uncertainties $<10$~Hz. 

\begin{figure}
    \centering
    \includegraphics[scale=0.6]{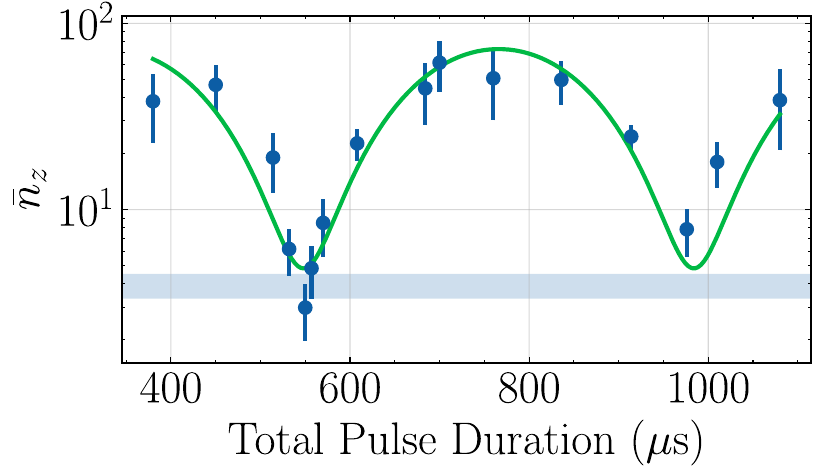}
    \caption{Measured axial mode occupation ($\bar{n}_z$) for various parametric exchange pulse durations coupling the axial and magnetron modes. Each experiment begins with Doppler and DR cooling of the axial degree of freedom, followed by application of a parametric axial-magnetron mode coupling excitation for a variable length of time. The axial mode occupation is then extracted from a 729~nm resonant carrier flop. The peak mode occupation measured corresponds to the original magnetron occupation ($\bar{n}_-$) before the parametric exchange. The vertical center of the transparent blue shaded region corresponds to the average occupation of the axial mode after Doppler and DR cooling before any exchanges. Its height corresponds to twice the standard error.}
    \label{fig:energyexchangevstime}
\end{figure}

In order to calibrate the mode coupling, first the axial mode is cooled with DR cooling, while the radial modes remain near their respective Doppler limits. Then, a mode coupling exchange pulse with the pre-calibrated sum or difference frequency of the modes is applied for a varied duration. The pulse coherently exchanges the mode occupations sinusoidally in time. We apply a resonant 729~nm laser pulse to probe axial sideband and carrier transitions to estimate the ion's axial temperature. At the appropriate pulse duration, the applied drive will induce a complete exchange of occupation between the selected modes, shown as the peaks in the final axial mode occupation of Fig.~\ref{fig:energyexchangevstime}. In Fig.~\ref{fig:energyexchangevstime}, the blue shaded region centered at $3.9(6)$ quanta is the DR cooling limit for these particular experiment parameters, and the total height of the shaded region is equal to twice the standard error. A sinusoidal fit to the blue data points is shown in green. To minimize off-resonant excitation of the spectator mode, the rf pulse is ramped on and off with $150~ \mu$s duration. The pulse time for a complete exchange of either mode is $\sim 350-400~\mu$s including $300~\mu$s total of pulse ramping with $\sim 1$~V amplitude applied to the segmented electrodes. The heating incurred on the axial mode during the exchange is determined by measuring the occupation of the mode after repeated applications of calibrated exchange pulses. A heating rate of $0.031(7)$ quanta per exchange is extracted after subtraction of the ambient axial heating during the exchange of $0.010(4)$~\footnote{The data of Fig.~\ref{fig:energyexchangevstime} reflects an elevated background axial mode heating rate due to technical noise. We have recently removed this technical noise source by electronically disconnecting the axialization inputs from the trap when not performing Doppler cooling.}. 

The motional mode exchange can be combined with DR cooling along a single mode direction (axial for this work) to sub-Doppler cool all degrees of freedom. First, the axial mode is cooled to the ground state, then the energy is swapped with the magnetron mode. The axial mode is then re-cooled which removes the energy that was initially in the magnetron mode. Next, the axial and modified cyclotron mode occupations are exchanged leaving minimal motional populations in the radial modes. Finally, the axial mode is re-cooled to the ground state. Notably, axial DR and sideband cooling cause recoil scatter into the two radial directions that are not actively being cooled. This recoil heating is currently non-negligible, and limits our final radial mode occupations to $\bar{n}_{\pm}\gtrsim10$.

\begin{figure}
    \centering
    \includegraphics[width=0.47\textwidth]{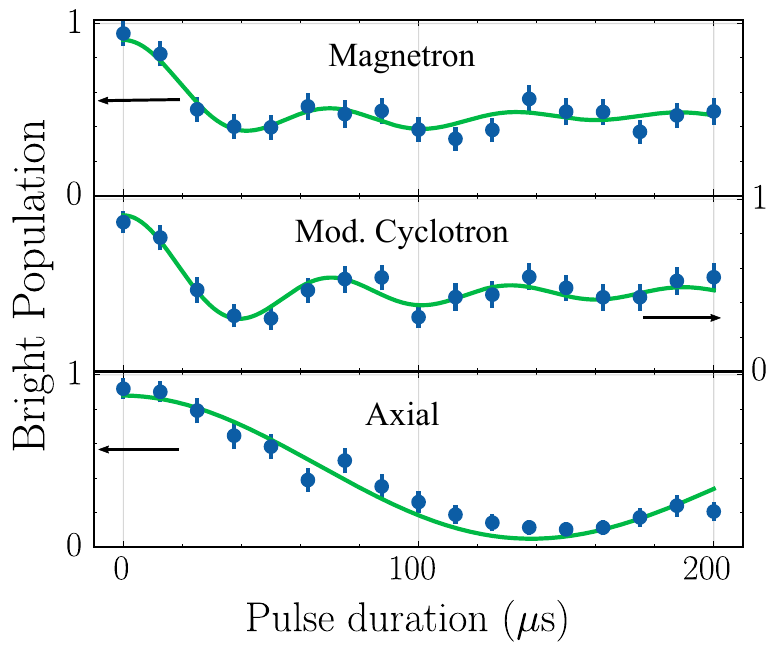}
    \caption{Plots of the bright populations while resonantly exciting the $S_{1/2},m_J=-1/2 \rightarrow D_{5/2},m_J=-3/2$ carrier (magnetron, modified cyclotron) or first blue sideband (axial) with an axially-propagating 729~nm laser beam. Each experiment consists of $800~\mu$s of DR cooling followed by 3~ms of pulsed sideband cooling. Sub-Doppler cooling of all modes is completed using only axially-propagating laser beams in the following sequence. First, axial motion is ground state cooled. Then, one of the radial mode occupations is exchanged with that of the axial and the axial cooling is repeated. This sequence is then repeated for the remaining radial mode. Finally, the axial occupation is measured via resonant 729~nm Rabi oscillations (following mode exchange pulses for the radial modes). The solid green lines are fits to the populations for extracting the final mode occupations of 21(4), 15(2), and 0.12(6) for the magnetron, modified cyclotron, and axial modes, respectively.}
    \label{fig:3dcooledmodes}
\end{figure}

Combining the techniques discussed above, the ion is cooled in three dimensions to sub-Doppler temperatures with DR laser cooling beams propagating only along a single axis. After $800~\mu$s of DR cooling followed by 3~ms of sideband cooling for each mode, the final thermal populations are measured via 729~nm sideband and carrier spectroscopy. The final mode occupations are $\bar{n}_- = 21(4)$, $\bar{n}_+ = 15(2)$, and $\bar{n}_z = 0.12(6)$, as obtained from the fits in Fig.~\ref{fig:3dcooledmodes}. Our semiclassical recoil heating estimates based on Eq.~\ref{eq:radial_dog} in Sec.~\ref{sec:model} yield an average radial mode heating rate of $\sim$14~ms$^{-1}$, suggesting final radial mode occupations of $\bar{n}_+=11$ and $\bar{n}_- = 22$ due to scattering from the DR cooling laser beams. The magnetron mode is the first to be exchanged with the axial, therefore it experiences two cycles of recoil heating while the modified cyclotron mode experiences one. Here, we neglect the relatively few recoil scatter events during the resolved sideband cooling steps.

\section{Semiclassical Model}\label{sec:model}
We model the DR cooling dynamics of \Ca~using a semiclassical treatment, where the internal state dynamics (i.e. photon scatter rates) are pre-computed via a Lindblad master equation formalism for a range of axial ion velocities. The external ion dynamics are subsequently evolved classically using the pre-computed and interpolated values for the photon scatter rate vs. ion velocity. 

Treating the axial sub-Doppler laser cooling with an effective optical wave vector ($k_{\mathrm{eff}}$) as one-dimensional, the force imparted on an ion is given by 
\begin{equation}
F_z = m \frac{dv_z}{dt}=-\hbar k_{\mathrm{eff}} \sum_i\Gamma_i \rho_{ii}(v_z),
\end{equation}
where $\hbar$ is Planck's constant, $v_z$ is the axial ion velocity, $m$ is the ion mass, $\Gamma_i$ is the inverse lifetime of atomic excited state $i$, and $\rho_{ii}$ is the velocity-dependent diagonal density matrix element (i.e. population) of excited state $i$. Integrating over all possible k-vectors for photon emission, the re-scattered photons will lead to diffusion of the total ion kinetic energy, $E$,  as~\cite{itano_laser_1982}
\begin{equation}
\frac{dE}{dt} = \sum_i {\frac{\hbar^2 k_r^2}{m} \Gamma_{i} \rho_{ii}},
\end{equation}
where $k_r$ is the wave vector for photon recoil. Combining the effects of photon absorption and re-emission, the resulting equation of motion for axial ion velocity is
\begin{eqnarray}\label{eq:EoM}
\frac{dv_z}{dt} &=& -\frac{\hbar k_{\mathrm{eff}}}{m} \sum_i \Gamma_i \frac{\partial \rho_{ii}}{\partial v_z}v_z + \sum_i\frac{\hbar^2 k_r^2}{3m^2} \Gamma_{i} \rho_{ii} \frac{1}{v_z}\nonumber \\
&\equiv& W(t) v_z + \frac{D(t)}{v_z},
\end{eqnarray}
where the first term of Eq.~\ref{eq:EoM} describes ion cooling (or heating) and the second term describes the energy diffusion in the $z$-direction due to spontaneous emission. Note that the form of $W(t)$ in Eq.~\ref{eq:EoM} includes a linear expansion of $\rho_{ii}(v_z)$. We neglect the Penning trap axial restoring force in Eq.~\ref{eq:EoM} since, in our system, the axial oscillation period of $\sim5~\mu$s is larger than the characteristic timescale for ion internal state dynamics ($\lesssim1~\mu$s). This is consistent with the weak binding limit discussed in Refs.~\cite{wineland_laser_1979,itano_laser_1982}.

For transition Doppler shifts that are large relative to the frequency width of the two-photon dark resonance feature, the $\rho_{ii}$ values will vary non-negligibly with time as the ion oscillates in the axial confining well. This effect is reflected in the explicit time dependence of the $W(t)$ and $D(t)$ terms. The k-vector for sub-Doppler cooling near the two-photon dark resonance is defined as $k_{\mathrm{eff}} \equiv \left|\vec{k}_{397} - \vec{k}_{866}\right|$, where $\vec{k}_{397}$ ($\vec{k}_{866}$) is the k-vector of the 397~nm (866~nm) laser beam. We apply the approximation $k_r = \left|\vec{k}_{397}\right|$ to describe recoil heating, which neglects the small fraction ($\sim8\%$) of spontaneous decays to the metastable $D_{3/2}$ manifold. 

For our sub-Doppler cooling with only axially-propagating laser beams, the radial degrees of freedom ($x$, $y$) experience only recoil heating. Applying the diffusion term of Eq.~\ref{eq:EoM}, we estimate the final radial velocity $\left( v_{r,f} = \sqrt{v^2_{x,f} + v^2_{y,f}}\right)$ after axial DRC for a duration, $t$, as
\begin{equation} \label{eq:radial_dog}
v_{r,f} = 2\sqrt{\int_0^{t}{D(t')dt'}},
\end{equation}
and we assume equipartition of the radial kinetic energy between the the modified cyclotron and magnetron modes.

As is evident in Eqs.~\ref{eq:EoM} and~\ref{eq:radial_dog}, computation of the cooling dynamics requires knowledge of the time-dependent atomic excited state populations. We calculate the excited state populations numerically following a Lindblad master equation treatment similar to that discussed in Refs.~\cite{allcock_dark-resonance_2016,Janacek_PhDThesis_2015}.

Our density matrix ($\rho$) for \Ca~in the Penning trap magnetic field includes all 18 magnetic sublevels within the $S_{1/2}$, $D_{3/2}$, $D_{5/2}$, $P_{1/2}$, and $P_{3/2}$ electronic states. The matrix $\rho$ therefore has dimension $18\times18$, and we solve for the 18-state populations and coherences using the Liouvillian superoperator as

\begin{equation}
\frac{d \vec{\rho}}{dt} = \textbf{L}\vec{\rho},
\end{equation}

where $\vec{\rho}$ is the vector form of $\rho$ with length $18^2=324$ and $\textbf{L}$ is the Liouville superoperator defined as

\begin{eqnarray} \label{eq:Liouville}
\textbf{L} &=& -i(\hat{H}\otimes \hat{I})+i(\hat{I}\otimes \hat{H}^T) \nonumber \\
&+& \sum_j{\Gamma_j \left[ 2\hat{\mathcal{J}}_j\otimes \hat{\mathcal{J}}_j^* - \hat{\mathcal{J}}_j^{\dagger}\hat{\mathcal{J}}_j\otimes \hat{I} - \hat{I}\otimes \hat{\mathcal{J}}_j^T\hat{\mathcal{J}}_j^* \right]}.
\end{eqnarray}

The first line of Eq.~\ref{eq:Liouville} includes the $18\times18$ system Hamiltonian $(\hat{H})$ and the identity $(\hat{I})$. The Hamiltonian describes the ion-laser electric dipole interactions under the rotating-wave approximation with the addition of the Doppler shift ($\propto -\vec{k}\cdot\vec{v}$) for a given ion velocity, $\vec{v}$. The second line of Eq.~\ref{eq:Liouville} describes each dissipative process, $j$, with the jump operator, $\hat{\mathcal{J}}_j$. Our dissipative processes consist of spontaneous state decay and laser frequency fluctuations, with the assumption of white laser frequency noise power spectral densities (i.e. Lorentzian laser linewidths). The jump operators for spontaneous decay from an upper state, $e$, to a lower state, $g$, take the form $\hat{\mathcal{J}} \propto |g\rangle\langle e|$. While we include the relatively small spontaneous decay rates of the metastable $D_{3/2}$ ($\sim 0.833$~s$^{-1}$) and $D_{5/2}$ ($\sim 0.856$~s$^{-1}$) levels, they have a negligible impact on the cooling dynamics discussed in this article. The primary decay mechanisms relevant for DR cooling in our setup are from the two $P_{1/2}$ sublevels ($\sim1.4\times10^8$~s$^{-1}$). 

We solve for the time-dependent axial velocity of the trapped ion in the regime of weak trap confinement. We therefore impose a harmonic oscillator condition on the velocity with the ansatz
\begin{eqnarray} \label{eq:harm_osc}
v_z(t) &=& \Re \left[A_v(t)e^{i(\omega_z t)} \right] \nonumber \\
&=& \Re \left[A_v(t)\right] \cos (\omega_z t) - \Im\left[A_v(t)\right] \sin (\omega_z t),
\end{eqnarray}
with time-varying velocity amplitude, $A_v(t)$. At each instance in time, we treat the ion as a free particle with instantaneous velocity defined by Eq.~\ref{eq:harm_osc}. Combining Eq.~\ref{eq:harm_osc} with Eq.~\ref{eq:EoM}, we produce the following differential equation for the complex velocity amplitude:
\begin{equation}\label{eq:amplitude}
\frac{dA_v}{dt} = \left[ W(t)-i\omega_z\right] A_v+\frac{D(t)}{A_v}e^{-2i\omega_z t}.
\end{equation}
The complex form of Eq.~\ref{eq:amplitude} avoids poles in the diffusion term that occur at the zero-velocity turning points of the ion oscillation during numerical integration. Previous numerical solutions of Ca$^+$ cooling relied on a modified diffusion term to mitigate this numerical instability~\cite{Janacek_PhDThesis_2015}.

To simulate the DR cooling results in Fig.~\ref{fig:drcvstime}, we numerically integrate Eq.~\ref{eq:amplitude} for a range of initial coherent state amplitudes, $|\alpha(0)|=\sqrt{mv_z^2(0)/(2\hbar\omega_z)}$, with uniformly randomized initial phases. The thermal trajectories of all initial excitation amplitudes $(0<|\alpha(0)|^2<3000)$ are plotted in Fig.~\ref{fig:drcvstime} as the light gray lines. To compute the mean thermal excitation, $\bar{n}_z$, vs. DR cooling duration (solid blue line of Fig.~\ref{fig:drcvstime}), we perform a thermally-weighted average of all simulated thermal trajectories, where the weights are consistent with the experimentally-measured initial thermal distribution with $\bar{n}_z=72$. 

We predict a finite DR cooling `capture range' of $|\alpha(0)|^2 < 900$ for these parameters, as evidenced by the heating trajectories for initial temperatures above this value in Fig.~\ref{fig:drcvstime}. For an initial thermal occupation of $\bar{n}_z=72$, this implies that approximately $1:230,000$ of DR cooling attempts will lead to axial heating. This effect is described as runaway heating in a recent publication reporting EIT cooling of $^{138}$Ba$^+$ outside the Lamb Dicke regime in an rf trap~\cite{bartolotta_laser_2024}. For extended DR cooling durations, our simulations predict that these low-probability heating trajectories will eventually increase $\bar{n}_z$ above the minimum values observed in the $\sim1$~ms timescale. Optimal DR cooling in the weak binding regime therefore requires maximizing this capture range without sacrificing cooling rate and minimum mode occupation, with the understanding that this minimum occupation may be transient in nature~\cite{lau_laser_2016}.

\section{Conclusion}
In summary, we have performed efficient axial ground state cooling in a Penning trap using the same laser beams required in the initial Doppler cooling step for DR cooling and performing a final sequence of pulsed sideband cooling with narrowband 729~nm excitations. Our total ground state cooling time of 3.8~ms represents a more than five-fold reduction compared with a previous sideband cooling demonstration with \Ca~ in the weak binding regime~\cite{hrmo_sideband_2019}. The measured DR cooling dynamics --- both the cooling rate and final mode occupation --- are well-described by a semiclassical model. We have also extended this cooling scheme to the radial modified cyclotron and magnetron modes using coherent mode exchange via trap potential modulation, ultimately achieving sub-Doppler mode occupations for all three eigenmodes.

The final temperatures of two of the three modes are limited by recoil heating from the DR cooling process. The magnitude of recoil heating varies with initial ion temperature, cooling duration, and the particular laser excitation parameters. We estimate that final recoil heating excitations below five quanta per uncooled mode are achievable with improved laser beam intensity stability and additional frequency stabilization of the 397A laser. It may also be possible to continuously couple an axial and radial degree of freedom \textit{during} the DR cooling process via the parametric mode exchange excitation. Our initial attempts along these lines have not been fruitful, but this is possibly due to uncompensated micromotion induced by the potential modulation modifying the DR cooling dynamics.   

The techniques described herein augment the growing set of laser cooling methods for quantum information experiments using atomic ions in Penning traps. Continued progress towards realization of a Penning trap quantum computing architecture will require efficient laser cooling schemes valid over a range of ion confinement conditions.

\begin{acknowledgments}
The authors thank Kenton R. Brown, Jonathan P. Home, and Daniel Kienzler for helpful discussions as well as Creston D. Herold for careful review of the manuscript. Research was sponsored by the Army Research Office and was accomplished under Grant Number W911NF-24-1-0355 The views and conclusions contained in this document are those of the authors and should not be interpreted as representing the official policies, either expressed or implied, of the Army Research Office or the U.S. Government. The U.S. Government is authorized to reproduce and distribute reprints for Government purposes notwithstanding any copyright notation herein.
\end{acknowledgments}

\bibliographystyle{apsrev4-2}
\bibliography{references}

\end{document}